\documentclass[onecolumn,noshowpacs,nofootinbib,colorlinks,hyperindex,unicode,10.5pt]{revtex4-2}
\usepackage{graphicx}
\usepackage{feynmp-auto}
\usepackage{tikz-feynman}
\usepackage{amssymb}
\usepackage{color,xcolor}

\usepackage{amsfonts,bm,amsmath,amssymb,tikz,accents}
\usepackage[eulergreek]{sansmath}
\usepackage{indentfirst,bigints}
\usepackage{hyperref,upgreek}
\usepackage{bm,graphics,color}
\usepackage{feynmp-auto}
\usepackage{slashed}
\counterwithout{equation}{section}
\usepackage{hyperref}
\usepackage{pgfplots}
\pgfplotsset{compat=1.16}

\hypersetup{hidelinks,colorlinks=true,breaklinks=true,urlcolor= blue}
\hypersetup{%
  colorlinks = true,
  linkcolor  = blue,
  citecolor = cyan,
}
\usepackage{textgreek}
\usepackage{color,xcolor}
\newcommand{\gdualn}[1]{\overset{\:{}^{{}^{\boldsymbol{\neg}}}}{\smash[t]{#1}}} 

\def\0{\mbox{\boldmath$\displaystyle\mathbb{O}$}}

\def\I{\openone}
\def\openone{\mathbb I}

\def\p{\mbox{\boldmath$\displaystyle\boldsymbol{p}$}}

\newcommand{\orcidicon}{%
	\begin{tikzpicture}
	\draw[lime, fill=lime] (0,0)
		circle [radius=0.16]
		node[white] {{\fontfamily{qag}\selectfont \tiny ID}};
	\draw[white, fill=white] (-0.0625,0.095)
		circle [radius=0.007];
	\end{tikzpicture}	\hspace{-2mm}
}
\newcommand\orcidg{{\href{https://orcid.org/0000-0002-7942-7941}{\orcidicon}}}
\newcommand\orcidRR{{\href{https://orcid.org/0000-0002-8283-2577}{\orcidicon}}}

\usepackage{float,xcolor,upgreek}
\usepackage{color} 
\usepackage{tikz-feynman}
\tikzfeynmanset{compat=1.0.0}
\newcommand{\beq}{\begin{eqnarray}}
\newcommand{\eeq}{\end{eqnarray}}
\newcommand{\bea}{\begin{eqnarray}}
\newcommand{\eea}{\end{eqnarray}}

\begin{document}

\title{Hermitian formulation for mass dimension one fermions:
Flat and curved space-times}

\author{Gabriel Brand\~ao de Gracia\orcidg{}}
\affiliation{Federal University of  Triângulo Mineiro, Physics Department, 
38064-200, Uberaba, MG, Brazil}
\email{gabriel.gracia@uftm.edu.br}

\author{Rodolfo Jos\'e Bueno Rogerio\orcidRR{}}
\affiliation{Centro Universitário UNIFAAT, Atibaia-SP, 12954-070, Brazil.}
\email{rodolforogerio@gmail.com}


\begin{abstract}
\indent  Throughout this paper, we conduct our discussion by a partial review of \cite{paper}, introducing the Hermitian formulation for interacting mass dimension one fermions based on Elko spinor. It includes pivotal observations about renormalizability and the study of some allowed interactions. Beyond these points, since dark-matter phenomenology is mainly connected to gravitation, we introduce original remarks on how the Hermitian prescription can be readily generalized to include curved space-time, considering the very definition of the Elko spinor structure. We establish the Elko dual as arising from the path-integral formulation of a more fundamental structure. It enables one to include a curved background space-time and also quantum gravity into our investigations.\\\\

\begin{center}
\noindent\textit{Dharam's rare blend of audacity, integrity, and brilliance reshaped how we think and aspire in science.}
\end{center}
\end{abstract}
\maketitle


\section{Introduction}

\indent Several experimental evidences points towards the existence of the so-called dark matter (DM) \cite{DM0,DM1,DM2,DM3,DM4}. It basically consists of measurements regarding gravitational lens \cite{glens1,glens2, glens3, glens4,glens5,glens6,glens7,glens8}, matter content in galaxy clusters \cite{cluster} and rotation curves \cite{curves}  of planar spiral galaxies. Moreover, one can mention the $\Lambda$CDM paradigm, the most successful cosmological model \cite{cdm}, based on a scenario with cosmological constant and cold (or non-relativistic) dark matter relic, in order to explain structure formation. It assumes that this non-luminous matter represents $\sim 25\% $ of the energy-matter content of the universe.\\
\indent Therefore, in analogy with ordinary matter,  one can envisage a field theory that describes the (DM) microscopic constituents. It then motivates a series of experimental initiatives based on the search of experimental signatures of ordinary matter/dark matter scattering cross sections such as Xenon-nT, LUX-ZEPLIN, and other approaches \cite{Persic:1995ru, Planck:2018vyg,XENON:2022ltv, Sassi:2022njl,LZ:2022lsv}. In this manner, mass dimension one fermionic fields based on Elko spinor, playing the role of expansion coefficient functions, enter as natural candidates for the description of this quite exotic kind of matter that mainly interacts through gravity, justifying the term dark \cite{elkodark1, elkodark2, elkodark3, elkodark4, elkodark5, elkodark6}. Interestingly, some intrinsic characteristics of our DM candidate allow a mechanism that implies highly suppressed luminous signs \cite{NPB2023}, in accordance with one of the most basic defining (DM) properties \cite{elkosmology1, elkosmology2, elkosmology3, elkosmology4, elkosmology5}.  Moreover, considering the demand for (DM) neutrality, one can easily verify that this spinor field formulation ensures that the external particles are eigenstates of the charge conjugation operator, reinforcing, once again, its dark nature. Although the external particles define eigenstates of this operator, this is not true for the quantum field as a whole. Then, there is a fermion number global U(1) charge. However, localizing it via minimal coupling prescription would lead to a non-Hermitian theory . Besides, the emergence of darkness is a direct consequence of two structurally connected aspects: $(i)$ the incompatibility in mass dimensionality between Standard Model fermionic fields and fermionic fields of mass dimension one inherently forbids the latter from participating in Standard Model $SU(2)$ doublet representations; $(ii)$ the underlying formalism governing mass dimension one fermions is intrinsically incompatible with the local gauge structure prescribed by the Standard Model. In other words, such fermions have renormalizable interactions with the SM particles only through gravity and an effective derivative interaction with the scalar part of the Higgs doublet arising after symmetry breaking, see the footnote on the introduction of \cite{paper}.  \\
 \indent Regarding some recent advances in the field, one can highlight a set of studies on the algebraic spinor properties \cite{spinorsdm1,spinorsdm2,spinorsdm3,Ahluwalia:2023slc,dharamnpb,rodo1,rodo2}, on the cosmological implications \cite{cheng,cosmo1,cosmo2} and also in the astrophysical context \cite{astrodm1,astrodm2,astrodm3,astrodm4}. 

This short review was developed in honor of the legacy of D.V. Ahluwalia, it is based on a series of recently published articles \cite{paper,dharamnpb,NPB2023}, which are associated with an effort to derive a rotationally covariant and a Hermitian formulation of mass-dimension one fields. Basically, we explore an analogy with the ordinary particle physics development establishing the basic fundamental properties, such as hermiticity, for obtaining a field theory capable of admitting a standard probabilistic interpretation, allowing the investigation of a series of scattering processes. A renormalizable coupling with the Higgs field is introduced in flat space as well as a possible curved space-time completion. The corresponding amplitudes for a set of physical processes are also analyzed. We emphasize that this study holds profound importance for the field, as it is based on a significant and recent advancement in the theory of Elko spinors and other mass-dimension one fields. The theory has now reached a more consolidated stage and aligns with the fundamental principles of quantum field theory.\\
\indent The paper is organized as follows: In sect. \ref{2}, we highlight the rotationaly covariant formulation and the adjoint structure. In sect. \ref{3}, we present the construction of the quantum field as well as a Hermitian interaction. The Sect. \ref{4} along with its subsections are devoted to present a set of Hermitian interactions, including the Feynman rules for the Elko-Higgs interaction, the renormalizability of such theory and we also present several amplitudes associated with different scattering processes involving mass-dimension-one fermions. In Sect. \ref{9} we discuss the path integral formulations that relates the second order formulation with a given first order one, using this approach to derive a possible route to a curved space-time completion of our Hermitian formulation. Then, in Sect. \ref{finalremarks} we conclude.

\section{On a rotationaly invariant basis and a prescription for the adjoint field}\label{2}
  
\indent One fundamental recent achievement of Elko research field is related to a field formulation compatible with the rotational covariance. Considering the anti-linear nature of the charge conjugation operator, one can derive a basis consisting of four particles and four anti-particles in the conjugate/anti-conjugate sets, respectively. This structure is capable of fulfilling the Weinberg rotational constraints for this exotic particle representation with additional degeneracies.

\indent Accordingly, we define the particle sector (the self-conjugate spinors)  \cite{dharamnpb}
\bea        \xi_1(\p)=\uplambda^{S}_{+}(\p)  \ ,\quad \ \xi_2(\p)=\uplambda^{S}_{-}(\p) \ ,\quad \ \xi_3(\p)=-i\uplambda^{A}_{+}(\p) \ ,\quad \ \xi_4(\p)=-i\uplambda^{A}_{-}(\p),  \label{basis}     \eea
and the anti particle sector (anti-self-conjugate spinors) 
     \bea        \upchi_1(p)=\uplambda^{A}_{+}(\p)  \ ,\quad \  \upchi_2(p)=\uplambda^{A}_{-}(\p) \ ,\quad \ \upchi_3(p)=-i\uplambda^{S}_{+}(\p) \ ,\quad \  \upchi_4(p)=-i\uplambda^{S}_{-}(\p),       \eea
     with both basis being defined in terms of the Elko spinors, as explicitly given in \cite{Ahluwalia:2023slc}.\\
\indent Taking into account the references \cite{wigner1,wigner2}, it is possible to assume a two-fold degeneracy in the presence of an anti-linear operator, doubling the degrees of freedom. This is an irreducible representation, since the rotational constraint demanded for one particle states can only be fulfilled by such a structure. The non-covariant parts are explicitly canceled in the spinor products.
 
Under the Dirac dual, Elko spinors have imaginary or vanishing norms \cite{jcap}. For this reason, and to fulfill the rotational constraint, the following adjoint structures are defined  \begin{align}
\gdualn{\xi}_{h}(\p) & = \left[ + \mathcal{P} \,\xi_{h}(\p) \right]^\dagger \gamma_0, \quad
\gdualn{\upchi}_{h}(\p)  =  \left[ - \mathcal{P} \,\upchi_{h}(\p) \right]^\dagger \gamma_0,
\end{align}
with $h=1,..,4$ standing for the spinorial label. As an additional comment, this adjoint formulation guarantees a set of allowed global symmetries as well as locality and a correct particle interpretation, see \cite{rodo1,rodo2}.  The parity operator $\mathcal{P}$ is defined in \cite{speranca} in the following fashion
\begin{equation}
\mathcal{P} \stackrel{\mathrm{def}}{=} m^{-1} \gamma_\mu p^\mu.
\end{equation}
\indent Explicit calculations yield the set of equations below, 
\begin{align}
\mathcal{P}\,\xi_{1}(\p) = +i \xi_{2}(\p), \quad  \mathcal{P}\,\xi_{2}(\p) 
= - i \xi_{1}(\p),\nonumber\\
\mathcal{P}\,\xi_{3}(\p)  = - i \xi_{4}(\p), \quad \mathcal{P}\,\xi_{4}(\p)=+i \xi_{3}(\p),
  \label{umm}\end{align}
and
\begin{align}
\mathcal{P}\,\upchi_{1}(\p) &=- i \upchi_{2}(\p), \quad \mathcal{P}\,\upchi_{2}(\p) =  + i \upchi_{1}(\p),\nonumber\\
\mathcal{P}\,\upchi_{3}(\p) & =  +i \upchi_{4}(\p),\quad \mathcal{P}\,\upchi_{4}(\p) = - i \upchi_{3}(\p),
\label{doiss}\end{align}
from which the full Elko spinor structure can be derived.

 Moreover, they also provide the additional fundamental information that all Elko spinor types are on-shell $p_\mu p^\mu=m^2$, ensuring a possible particle description.\\
Considering the adjoints thus defined, one can obtain the orthonormal relations
\begin{equation}
\gdualn{\xi}_{h}(\p) {\xi}_{h^{\prime}}(\p) = 2 m \,\delta_{h{h}^\prime},\; \
\gdualn{\upchi}_{h}(\p) {\upchi}_{h^{\prime}}(\p) = - 2 m \,\delta_{h{h}^\prime},\nonumber
\end{equation}
and the spin sums being clearly Lorentz invariant 
 \begin{align}
  \sum_{h} \xi_{h}(\p) \gdualn{\xi}_{h}(\p)  = 2m\, \I_4, \; \
  \sum_{h} \upchi_{h}(\p) \gdualn{\upchi}_{h}(\p)  =
 -  2m  \,\I_4 .
 \end{align}
Such important features are only possible due to the well-defined adjoint structure. The study of dual structures constitutes a theory in its own right; significant details regarding recent developments and support for mathematical adjoint definition can be found in \cite{rodolfohidden, rodolfotaka, juliorogerio}. 
 
\section{On action formulation, interactions, and the Hermitian nature} \label{3}

\indent From now on, in order to model dark matter particles, we consider the following quantum field based on the Elko spinor playing the role of expansion coefficients functions. The free field structure is given by
\bea \uplambda(x)=\int \frac{d^3p}{(2\pi)^3\sqrt{2mE(p)}}\left[ \left(\sum_{h=1}^{4}c_h(\p)\xi_h(\p)  \right)e^{-ip.x}+\left(\sum_{h=1}^{4}d^\dagger_h(\p)\upchi_h(\p)  \right)e^{ip.x}   \right],       \label{field1}     \eea
whose dual reads 
\bea \gdualn{\uplambda}(x)=\int \frac{d^3p}{(2\pi)^3\sqrt{2mE(p)}}\left[ \left(\sum_{h=1}^{4}c^\dagger_h(\p)(\mathcal{P}\xi_h(\p))^\dagger  \right)e^{ip.x}+\left(\sum_{h=1}^{4}d_h(\p)(-\mathcal{P}\upchi_h(\p))^\dagger  \right)e^{-ip.x}   \right] \gamma_0  \label{field2}        \eea
As an important observation, considering the structure of the parity operator, the dual field can be expressed as 
\bea\label{gdd} \gdualn{\uplambda}(x)=\left(i\frac{\slashed{\partial}}{m} \uplambda(x)\right)^\dagger \gamma_0.             \eea
As mentioned, the spinor coefficients related to the particle and anti-particle sectors are, respectively, self-conjugate and anti-self-conjugate under the action of the charge conjugation operator.\\ 
\indent It is important to note that for the case of spin one half bosons \cite{Ahluwalia:2023slc,dharamboson}, the adjoint of both particle and anti-particles have the same prescription. The lack of an additional minus sign (in the dual structure definition) prevents one from deriving the full adjoint formulation above as well as the correlated Hermitian structures that are going to be studied here. This is a result in agreement with the spin-statistics theorem \cite{spin}.\\
Now, considering Eq. (\ref{gdd}), one notices that the action below, defining the on-shell nature of the free field, is Hermitian, up to a non-contributing boundary term, 
\bea \mathcal{S}=\int d^4x \left[\partial_\mu \gdualn{\uplambda}(x)\partial^\mu\uplambda(x)-m^2\gdualn\uplambda(x)\uplambda(x)              \right].\label{lagr} \eea
It is worth mentioning that the propagator arising from the inverse of the differential operator of the model indeed equals the time ordered product of Elko fields expressed in terms of the previously introduced field expansion \cite{rodo2}. This is an important consistency check to ensure compatibility with the optical theorem \cite{opticaltheo}.\\
\indent In this manner, rewriting the scalar bilinear as  \bea \gdualn{\uplambda}(x)\uplambda(x)=-i\uplambda^\dagger(x)\gamma_0 \Big(\frac{\gamma^\mu \stackrel{\leftarrow}{\partial_\mu}}{m}\Big)^{ \dagger}\uplambda(x),     \label{herm1}\eea
  using the identity $\frac{\slashed{\partial}^\dagger}{m}\gamma_0=\gamma_0\frac{\slashed{\partial}}{m}$, and discarding a total derivative term, it immediately leads to
\bea \mathcal{S}=\mathcal{S}^\dagger. \label{herm2}\eea

A Hermitian and renormalizable interaction with the Higgs boson can be described by the following Lagrangian term
\bea \mathcal{L}_I=\gdualn{\uplambda}(x)i\slashed{\partial}\uplambda(x)\phi(x)g'. \eea 
It is straightforward to show that the interaction Lagrangian itself is fully Hermitian, since
\begin{eqnarray}
    \Big[\frac{1}{m}(i\slashed{\partial}\uplambda(x))^\dagger\gamma_0(i\slashed{\partial}\uplambda(x))\phi(x) g'    \Big]^\dagger&=&\Big[\frac{1}{m}(i\slashed{\partial}\uplambda(x))^\dagger\gamma_0^\dagger \Big((i\slashed{\partial}\uplambda(x))^\dagger\Big)^\dagger \phi(x) g'    \Big]\nonumber\\&=&\gdualn{\uplambda}(x)i\slashed{\partial}\uplambda(x)\phi(x)g', \end{eqnarray} 
    where $g'$ is a dimensionless coupling constant. We highlight that equations \eqref{herm1} and \eqref{herm2} hold profound significance within Elko theory, furnishing new perspectives for advancements across numerous fields.
    
\section{Hermitian interactions and a criterion for a dark coupling}\label{4}
\indent The previous definitions make possible the derive of a set of Hermitian interactions suitable to, at least, define effective models. The first one consists of a
Yukawa-like interaction with the Higgs boson, given by \cite{paper}
\bea \mathcal{L}_I=g'\gdualn{\uplambda}(x)i\slashed{\partial}\uplambda(x)\phi(x).  \eea 
It furnishes amplitudes whose structures are different from all the known ones from the standard model. It is renormalizable and Hermitian.\\
\indent One can also define a Hermitian extension of the neutral Pauli-like coupling with the electromagnetic field, as
\bea  \mathcal{L}_I=\tilde{g}\gdualn{\uplambda}(x)i\left[\gamma^\mu,\gamma^\nu \right]F_{\mu \nu}(i\slashed{\partial}\uplambda(x)),                 \eea
with $\tilde{g}$ being a coupling with a dimension of inverse powers of mass.
We emphasize that the previous pseudo-Hermitian version is associated with a darkness mechanism regarding the coupling with mass dimension one fermions constructed from Elko spinors  \cite{NPB2023}.

The Elko effective coupling with Dirac fermions such as the neutrino,  can be deformed to become Hermitian as
\bea  \mathcal{L}_I=\mathrm{c}_1\gdualn{\uplambda}(x)(i\slashed{\partial})\uplambda(x)\Bar \eta(x) \eta(x),                 \eea
with $\mathrm{c}_1$ being another dimensionful coupling parameter as well as $\mathrm{c}_2$ and $\mathrm{c}_3$ present in the next two interactions highlighted in this section.

The  Elko-Higgs coupling with four legs is Hermitianized as
\bea   \mathcal{L}_I=\mathrm{c}_2\gdualn{\uplambda}(x)(i\slashed{\partial})\uplambda(x)\phi^2(x),                \eea 
 whereas the  self-interaction can have a Hermitian version like
\bea    \mathcal{L}_I=\mathrm{c}_3\Big(\gdualn{\uplambda}(x)(i\slashed{\partial})\uplambda(x) \Big)^2.          
   \eea
\noindent Therefore, although these approaches define legitimate models by themselves, we consider that the final theory must undergo renormalizability and Hermiticity, defining our physical requirements. This criterion fixes the Yukawa-like coupling as a viable portal for investigating DM physics. As we will see, this interaction is suitable to explain some of the observed experimental data in a natural setting. Finally, if one wants to transform the global fermion number U($1$) invariance into a local symmetry through introduction of a gauge field via minimal coupling, the associated model would be non-Hermitian and ruled out by this physical criterion. It reinforces the dark nature of the Elko field. \\

\subsection{Elko-Higgs derivative coupling}\label{5}

\begin{figure}[H]
\centering
\includegraphics[width=13.15cm] {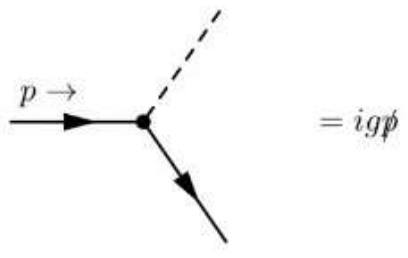}  \vspace*{-7.5cm}
\caption{Elko-Higgs coupling. The arrows denote mass dimension one fermions and the dashed line refers to the Higgs particle.}
\label{full}
\end{figure}

\indent Taking into account Hermiticity and renormalizability, we consider the derivative Yukawa theory as the main paradigm. Hence, the following action is considered
\bea \!\!\!\!\!\!\!\!\!\!  S\!=\!\!\int\! d^4x\Bigg[\partial_\mu \gdualn{\uplambda}(x)\partial^\mu\uplambda(x)\!-\!m^2\gdualn\uplambda(x)\uplambda(x)\!+\!\frac{1}{2}\big(\partial_\mu \phi \partial^\mu \phi\!-\!M^2\phi^2\big) \!+\!\gdualn{\uplambda}(x)i\slashed{\partial}\uplambda(x)\phi(x)g'\!+\!\cdots  \Bigg], \eea
where the symbol $\cdots$ stands for the remaining Higgs \footnote{Here, we consider the neutral part of the Higgs doublet after spontaneous symmetry breaking.} couplings in the electroweak sector as well as its quartic self-interaction.\\
\indent The interaction is associated with the vertex shown in Fig. \ref{full}.  
    \vspace*{-8cm}
      
\begin{figure}[H]
\centering
\includegraphics[width=14cm]{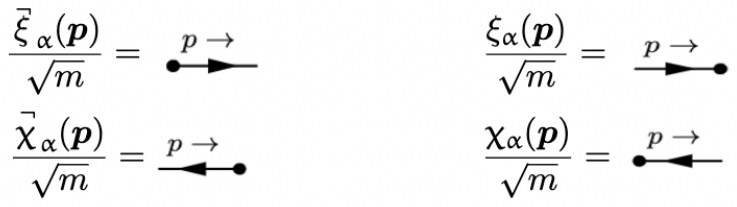} \vspace*{-8.5cm}
\caption{Feynman rules for external Elko fields.}
\label{full2}
\end{figure}

       The Feynman rules for external Elko fields are depicted in Fig. \ref{full2}. \footnote{The $h$ label is replaced by the $\alpha$ one. This is just a notational change convenient for the next discussions.}
   \vspace*{-8.5cm}

The mass dimension one fermion associated with DM has the following propagator 
\bea \frac{i}{k^2-m^2+i\epsilon},\eea
whereas the Higgs propagator reads
\bea  \frac{i}{k^2-M^2+i\epsilon}.\eea
\noindent Therefore, from these building blocks, one can derive the 1-loop radiative corrections and explicitly discuss renormalizability. Later, a set of scattering processes can be analyzed,  to define the role of mass dimension one fermions constructed in terms of Elko spinors in the context of the DM phenomenology. The Feynman rules also reveal another important aspect of theories based on the Elko spinors: Although the whole mass dimension one quantum field is not invariant under charge conjugation, the external states indeed define its eigenstates. 

\subsection{Renormalizability at one loop}\label{6}

This section is devoted to the discussion of renormalizability. We explicitly compute the divergent $n$-point functions at 1-loop approximation and show that the singular pieces have the same form as the terms originally present in the bare Lagrangian. This means that they can be absorbed in a consistent renormalization scheme such as that in the on-shell one, for example. Although power-counting arguments play an important role in quantum field theory, there are cases in which even a power-counting renormalizable theory may need ultraviolet completion to ensure finiteness. We can mention the illustrative case of quantum scalar electrodynamics \cite{rohr}, in which it is necessary to add an additional self-interaction term for the scalar particles, despite the good power-counting arguments for this specific interaction. Therefore, a careful explicit evaluation of the 1-loop renormalized structure is required.\\
\indent Considering the Feynman parametrization, the bosonic self-energy reads 
\bea\label{divvf}
 i\Pi_\phi(p)= {g'}^2\int \frac{d^4k}{(2\pi)^4}\frac{Tr[(\slashed{p}-\slashed{k})\slashed{k}]}{[(p\!-\!k)^2\!-\!m^2][p^2\!-\!m^2]}={g'}^2\!\!\int \!\frac{d^4k}{(2\pi)^4}\int_0^1\! dx\frac{Tr[p^2x(1\!-\!x)\!-\!k^2]}{[k^2-\Delta_x]^2},\eea
with $\Delta_x=m^2-p^2x(1-x)$. The divergent part of Eq. \eqref{divvf} reads \footnote{Considering dimensional regularization in the limit $\Bar \epsilon \to 0$. }
\bea \left(\Pi_\phi(p)\right)_{div}= \frac{{g'}^2}{16\pi^2\Bar \epsilon}\,p^2-m^2 \frac{{g'}^2}{8\pi^2\Bar \epsilon},         \eea
which has indeed the same type of the bare Lagrangian.\\
\indent The fermionic self-energy is given by
\bea i\Sigma(p)={g'}^2\int \frac{d^4k}{(2\pi)^4}\frac{\slashed{k}\slashed{p}}{(k^2-m^2)\left((k-p)^2-M^2\right)}={g'}^2p^2\int \int_0^1dx  \frac{d^4k}{(2\pi)^4}\frac{x}{\left(k^2-\Delta'_x\right)^2},                      \eea for 
$\Delta'_x=(1-x)(m^2-p^2x)+xM^2$. The divergent piece has the same form as the  Gaussian kinetic term in the bare Lagrangian and reads
\bea    \Big(\Sigma(p)\Big)_{div}=\frac{{g'}^2}{16\pi^2\Bar \epsilon}\,p^2,   \eea
ensuring renormalizability. 
Hence, the remaining divergent radiative correction is the vertex function
\bea  i\Gamma(q_1,p)= {g'}^3\int \frac{d^4k}{(2\pi)^4}\frac{[(\slashed{p}+\slashed{k}) \slashed{k} \slashed{q}_1]}{[(k+q_1)^2-M^2][k^2-m^2 ][(p+k)^2-m^2 ]}, \eea
with $q_1$ being the fermionic momentum entering the graph in the same arrow orientation. It can be parametrized as
\bea     [i\Gamma(q_1,p)]_{div}=\slashed{q}_1 {g'}^3\int \int dx\,dy\,dz \,\delta(x+y+z-1)\frac{d^4k}{(2\pi)^4}\frac{k^2}{[k^2-\tilde{\Delta}_x]^3}, \eea                              with $\tilde{\Delta}_x=-xyp^2+(1-z)^2m^2$. Its divergent piece has also the same form as bare Lagrangian
\bea     [\Gamma(q_1,p)]_{div}=\slashed{q}_1 \frac{{g'}^3}{16\pi^2  \Bar \epsilon},  \eea  
completing the leading-order renormalizability verification.\\
\indent According to the next developments, it is possible to define the model's parameter configurations in compliance with experimental data in which $g'$ is not prohibitively small, meaning that the study of the radiative corrections is relevant for this phase.\\
\indent Then, summarizing, the presence of the derivative interaction adds a power of the integration momentum in each fermion internal line. Moreover, considering the specific structure of the DM propagator and the fact that the trilinear graph topology is the same as the standard QED$_4$ one, it is possible to show that the superficial degree of divergence resembles the expression concerning QED$_4$ \footnote{Here, $N_\phi$ and $N_\uplambda$ denote the number of external Higgs fields and the number of external Elko, respectively.},
\bea D=4-N_\phi-\frac{3}{2}N_\uplambda,    \eea
meaning that the four-point 1PI function with external bosonic lines is logarithmic-divergent. Differently from  QED case, there is no symmetry implying the decrement of the divergence degree for this specific radiative correction. Fortunately, the divergent piece is constant and can be absorbed in a renormalization procedure involving the bare Higgs four legs vertex already present in standard model, completing our 1-loop analysis. Finally, according to previous claims, although our Yukawa-like interaction with the Higgs self-interaction completion seems to be fully renormalizable, one must also consider the other interactions in electroweak sector in which the Higgs takes part. The relation between these different kinds of interactions certainly leads to a more compicate analysis.

\subsection{M\o ller-like scattering}\label{7}

\indent   The Hermitian interaction with the Higgs boson leads to an interesting phenomenology emerging from the wide set of electroweak couplings involving this scalar particle field.
Regarding these processes, it is worth mentioning that the overall topology of the graphs is trilinear, a well-known structure, and the processes are prototypical processes of QFT (the channels are well-known); the only changes rely on a specific functional form of the amplitudes. Moreover, the specific ELKO external lines are also present in the paper, ensuring the fulcrum tools to build the amplitudes.\\
\indent In order to start this discussion, one can evaluate the M\o ller-like scattering between Elko particles mediated by the Higgs. The appropriate scattering parametrization in the center-of-mass frame reads
\begin{subequations}
\bea    p_1^\mu&=&(E,0,0,p), \qquad\qquad\qquad\qquad p_2^\mu=(E,0,0,-p),\\
p_3^\mu&=&(E, p \sin \theta,0, p \cos\theta), \qquad\qquad p_4^\mu=(E, -p \sin \theta, 0, -p \cos\theta),              \eea
\end{subequations}
with $E=\sqrt{p^2+m^2}$. This scattering process has contributions from the $t$ and $u$ channels with the following associated amplitude
\bea
    i{\cal{M}}_{\upalpha, \upbeta, \upbeta^\prime, \upalpha^\prime}&=&\frac{{g'}^2}{m^2\big (t-M^2\big)}i\gdualn  
 \xi_\upalpha(\p_4)\slashed{p}_2\xi_\upbeta(\p_2) \gdualn \xi_{\upbeta'}(\p_3)\slashed{p}_1\xi_{\upalpha'}(\p_1)\nonumber \\
    &&-\frac{{g'}^2}{m^2 \big(u-M^2\big)}i\gdualn \xi_{\upbeta'}(\p_3)\slashed{p}_2\xi_{\upbeta}(\p_2) \gdualn \xi_{\upalpha}(\p_4)\slashed{p}_1\xi_{\upalpha'}(\p_1),
\eea
expressed in terms of the Elko spinor types/labels. Interestingly, it leads not only to positive squared amplitude taken with the usual $\dagger$ prescrition but also guarantees a well-defined probabilistic interpretation. It enables one to do physics with Elko spinors and compare results with experiments such as Xenon-nT \cite{XENON:2022ltv}.\\
\indent Regarding the squared polarized amplitudes, we can mention some of them like \footnote{They are explicit positive definite, reflecting the Hermitian nature.} 
\begin{eqnarray}\label{ampli111}
 |{\cal{M}}_{1,1,1,1}|^2 = \left(
\begin{array}{c}
 \frac{{g'}^4  p^4 \sin ^4(\theta ) (m+2 E )^4 (p^2-E^2)^2 (t-u)^2 \left((m+E)^2-p^2\right)^2 \left(-M^2+t+u\right)^2}{16 m^{12}  (m+E )^4 \left(M^2-t\right)^2 \left(M^2-u\right)^2} \\
\end{array}
\right),   
\end{eqnarray}
highlighting the fact that the very same result is obtained for the cases $|{\cal{M}}_{2,2,2,2}|^2$, $|{\cal{M}}_{3,3,3,3}|^2$, and $|{\cal{M}}_{4,4,4,4}|^2$.

One can also mention other possible transitions such as
\begin{eqnarray}\label{ampli1122}
 &&|{\cal{M}}_{1,1,2,2}|^2 =\left(
\begin{array}{c}
 \frac{{g'}^4 p^4 \sin ^4(\theta ) (m+2 E )^4 (p+E )^4 (t-u)^2 (m+p+E )^4 \left(-M^2+t+u\right)^2}{16 m^{12}   (m+E )^4 \left(M^2-t\right)^2 \left(M^2-u\right)^2} \\
\end{array}
\right),
\end{eqnarray}
and 
\begin{eqnarray}\label{ampli3434}
  |{\cal{M}}_{3,4,3,4}|^2 =  
\begin{array}{c}
 \frac{{g'}^4  (p^2-E^2 )^2  (t-u)^2 \big((m+E)^2-p^2 \big)^2  \left(-M^2+t+u\right)^2 \left[-\left(p \cos (\theta ) (m+2 E )\right)^2+\left(m (m+E )+2 p^2\right)^2\right]^2}{16 m^{12}  (m+E )^4 \left(M^2-t\right)^2 \left(M^2-u\right)^2} \\
\end{array}.
\end{eqnarray}

\subsection{On the Elko annihilation processes}\label{8}

\indent The study of Elko annihilation processes is a useful tool to constrain the model's parameters through cosmological requirements associated with the current measurements of DM abundance. As we shall see, the coupling with the Higgs model ensures a rich DM phenomenology that complies with experimental data. \\
\indent Regarding the Elko annihilation processes, the situation is suitably parameterized by the variables
\begin{subequations}
\bea    p_1^\mu&=&(E, 0, 0, p), \qquad\qquad\qquad\qquad p_2^\mu=(E, 0, 0, -p),\\
p_3^\mu&=&(E, p' \sin \theta, 0, p' \cos\theta), \qquad\qquad p_4^\mu=(E, -p' \sin \theta, 0, -p' \cos\theta),              \eea
\end{subequations}
with $p_1^\mu$ and $p_2^\mu $ representing the 4-momentum of the incoming DM particles in the center-of-mass frame. The remaining variables denote the 4-momentum of the outgoing produced particles. For the case of an annihilation process of two incoming Elko into two outgoing Higgs particles, one can set the energy as $E=\sqrt{p^2+m^2}=\sqrt{p^{'2}+M^2}$, with $m$ and $M\approx 125$ GeV representing the Elko spinor and Higgs masses, respectively. \\
\indent Considering the Feynman rules, the amplitude reads
\bea i{\cal{M}}_{\beta \sigma}=i\frac{{g'}^2}{m(t-m^2)} \gdualn{\upchi}_\beta(\p_2)(\slashed{p}_1-\slashed{p}_3)\slashed{p}_1 \xi_\sigma(\p_1)+i\frac{{g'}^2}{m(u-m^2)}\gdualn{\upchi}_\beta(\p_2)(\slashed{p}_1-\slashed{p}_4)\slashed{p}_1 \xi_\sigma(\p_1).                     \eea

\indent Since the Higgs boson has Yukawa-like couplings with a variety of particles, one should also consider $s$-channel annihilation processes into a pair of fermions due to the electroweak couplings of the form $g_{\mathrm{J}}\phi(x) \Bar\psi_{\mathrm{J}}(x) \psi_{\mathrm{J}}(x)$, in which the label ${\mathrm{J}}$ regards the fermionic types ${\mathrm{J}}=e,u,d$  associated with the electron, the up, and the down quarks, respectively. The amplitude of the process reads \footnote{Here, $\Bar u_{\gamma'}^\mathrm{J}(p_3)$ and $v_{\sigma'}^\mathrm{J}(p_4)$ denote the electron and positron spinors, respectively. The label $\gamma$ and $\sigma$ are associated with the spin.}
\bea i{\cal{M}}_{\alpha \beta \gamma' \sigma' }=-i\frac{g'g_{\mathrm{J}}}{m(s-M^2)} \gdualn{\upchi}_\alpha(\p_1)\slashed{p}_2 \xi_\beta(\p_2) \Bar u_{\gamma'}^{\mathrm{J}}(p_3) v_{\sigma'}^{\mathrm{J}}(p_4).                   \eea
\indent The non-polarized squared amplitudes have the following expression
\bea \frac{1}{8}\sum_{\alpha \beta \gamma' \sigma' }|{\cal{M}}_{\alpha \beta \gamma' \sigma' }|^2=\frac{(g'g_{\mathrm{J}})^2}{8(sM^2)^2}Tr\left[\slashed{p}_1\slashed{p}_2\right]Tr\Big[(\slashed{p}_3+m_\mathrm{J})(\slashed{p}_4-m_\mathrm{J})\Big],            \eea
with $m_e\approx 0.5$ MeV,  $m_u\approx 2.15$ MeV and $m_d\approx 4.5$ MeV. The coupling constants have magnitudes $g_e\approx 2\times 10^{-6}$, $g_u\approx  10^{-5}$,  and $g_d\approx 2\times 10^{-5}$. It is worth mentioning that according to further developments in the next sections, this weak coupling is the origin of the difficulty in observing DM/electron scattering in the laboratory, which means that the investigation of this portal leads to a natural description of such phenomenology.\\
\indent Beyond the $s$-channel annihilation into fermion pairs, one can also consider this kind of annihilation involving vector bosons. They are associated with the electroweak Higgs couplings of the form \footnote{The symbols $V^\mu_A(x)$ denote the vector fields representing the $Z$ and $W$ vector bosons. }
\bea g_A m_A\phi(x)\eta_{\mu \nu}|V^\mu_A(x)| |V^\nu_A(x)|,\eea with $A=Z,W_+,W_-$ recovering the cases of the weak $Z$ and $W$ vector bosons, respectively. The amplitude reads \footnote{Here $\epsilon_\mu^r$ represent the polarization vectors of the vector boson fields.}
\bea i{\cal{M}}_{\alpha, \beta r,s}=-im_A\frac{g'g_A}{m(s-M^2)} \gdualn{\upchi}_\alpha(\p_1)\slashed{p}_2 \xi_\beta(\p_2)\eta^{\mu \nu}\epsilon_\mu^r\epsilon_\nu^s.                    \eea
The averaged non-polarized amplitude for these massive bosons
\footnote{ Considering the relation $\sum_r\epsilon_\mu^r \epsilon_\nu^{*r}=-\eta_{\mu \nu}+\frac{p_\mu p_\nu}{m^2_A}$ for the polarization vectors.}
\bea \frac{1}{12}\sum_{\alpha \beta r s }|{\cal{M}}_{\alpha \beta r s }|^2=\frac{m^2_A(g'g_A)^2}{4(s-M^2)^2}Tr\left[\slashed{p}_1\slashed{p}_2\right],     \eea
with $m_Z\approx 91.19$ GeV and $m_W\approx 80.38$ GeV, whereas for the  couplings one has $g_z\approx 0.36$ and $g_w\approx 0.32 $.\\

\section{Elko action from a more fundamental structure and a possible curved space-time completion}\label{9}

\indent In order to discuss some fundamental aspects as well as a possible coupling with gravity, since DM is inferred by gravitational phenomenology, it is important to highlight an equivalent first-order formulation in terms of the Dirac dual. It is based on two fields, $\uplambda$ and $\tilde{\uplambda}$ that are related only in the mass shell, with one becoming the parity reversed version of the other. It is defined in the first line of the equation below\footnote{The Dirac dual $\uplambda^\dagger(x) \gamma_0$ is denoted as $\bar{\uplambda}(x)$.}

 \begin{eqnarray} Z&=&N\int \mathcal{D}\phi  \, \mathcal{D}{\bar{\uplambda}}\, \mathcal{D} {\uplambda}\mathcal{D}{\bar{\tilde \uplambda}}\, \mathcal{D} {\tilde \uplambda}\exp\left[i\int d^4x\left( im\bar \uplambda \slashed{\partial}\uplambda+im\bar {\tilde{\uplambda}} \slashed{\partial}\tilde{\uplambda}-m^2\bar {\tilde{\uplambda}} {\uplambda}-m^2\bar {{\uplambda}} \tilde{\uplambda} +m\bar {\tilde{\uplambda}} \tilde{\uplambda}\phi g'
  \right)\right]\nonumber \\  \ && =N\int \mathcal{D}\phi  \, \mathcal{D}{\bar{\tilde \uplambda}}\, \mathcal{D} {\uplambda}\mathcal{D}{{\tilde \uplambda}}\, \delta(i\slashed{\partial}\uplambda-m\tilde \uplambda)\nonumber \exp\left[i\int d^4x\left( +im\bar {\tilde{\uplambda}} \slashed{\partial}\tilde{\uplambda}-m^2\bar {\tilde{\uplambda}} {\uplambda} +m\bar {\tilde{\uplambda}} \tilde{\uplambda}\phi g'
  \right)\right]\nonumber \\  \ && =N'\int \mathcal{D}\phi  \, \mathcal{D}{\gdualn{\uplambda}}\, \mathcal{D} {\uplambda}\exp\left[i\int d^4x\left( -\gdualn{\uplambda}(\Box+m^2)\uplambda+\gdualn{\uplambda} i\slashed{\partial}{\uplambda}\phi g'
  \right)\right].  \label{path}        \end{eqnarray}

  \indent Interestingly, assuming both fields to be independent, the model described in the first line above is such that the equations\eqref{umm} and \eqref{doiss}, defining the whole Elko structure,  arise from the minimal of the action in the limit $g' \to 0$, as can be explicitly verified in the appendix of \cite{Ahluwalia:2023slc}. It ensures the obtainment of the Elko fields present in the definition of the quantum field given in \eqref{field1} and \eqref{field2} as well as their parity reversed versions. After functional integration, the effective action for the mass dimension one fermions is recovered. In this case, the information regarding the Elko structure (beyond just the on-shell nature) is implicit in the adjoint prescription. More specifically, the Elko set has zero norm under Dirac prescription, although there are non-vanishing crossed products with their parity reversed versions. This is the fundamental content of the adjoint prescription. One can also notice that the first and second-order formulations are also equivalent in the thermodynamical point of view, since the path integrations relating both structures do not lead to operator-dependent overall coefficients. Finally, it is interesting to note that the resulting theory obtained after integration would be the same if one replaces the original first-order model by a theory characterized by interchanging the roles of the $\uplambda(x)$ and $\tilde{\uplambda}(x)$ fields. This symmetry can possibly be related to the underlying Wigner degeneracy.  \\
  \indent According to \cite{daRocha:2011yr}, the curved space-time generalization of the equations \eqref{umm} and \eqref{doiss}, defining the Elko in flat space is obtained from the general covariance requirement $\slashed{\partial} \to \slashed{\nabla}$. Therefore, this set of equations implies 
  \bea (\slashed{\nabla}\slashed{\nabla}+m^2)\uplambda=0,\eea based on the Lichnerowicz operator being essentially different from $(\nabla_\mu \nabla^\mu +m^2)\uplambda=0$, as correctly pointed out by \cite{cheng}.\\
  \indent Therefore, considering for a while the metric just as a space-time function and not a quantum field, one can replace $\slashed{\partial} \to \slashed{\nabla}$ e $d^4x \to \sqrt{-g}d^4x$ in the path integral \eqref{path} leading, after integration, to
\bea   Z=  N'\int \mathcal{D}\phi  \, \mathcal{D}{\gdualn{\uplambda}}\, \mathcal{D} {\uplambda}\exp\left[i\int \sqrt{-g} d^4x\left( -\gdualn{\uplambda}(\slashed{\nabla}\slashed{\nabla}+m^2)\uplambda+\gdualn{\uplambda} i\slashed{\nabla}{\uplambda}\phi g'
  \right)\right],           \label{46}                    \eea
with the adjoint prescription in a covariant fashion $\gdualn{\uplambda}=\frac{1}{m}(i\slashed{\nabla}\uplambda)^\dagger\gamma_0$. Considering this definition, and the insightful reference \cite{shapiro}, we have the covariant derivative structure
\bea \nabla_\mu \uplambda(x) \equiv \partial_\mu \uplambda(x)+\frac{i}{2}\omega_\mu^{ab}(x)\sigma_{ab}\uplambda(x), \eea in terms of the spin connection relating local Lorentz (LL) frames and generators $\sigma^{ab}=\frac{i}{2}(\gamma^a\gamma^b-\gamma^b\gamma^a)$. Also, the Dirac gamma matrices $\gamma_\mu$ associated with general coordinate transformations are related to the (LL) structures $e^\mu_a(x)\gamma^a=\gamma^\mu$ through the tetrad. Due to the assumed metric compatibility, we consider the tetrads to also be covariantly conserved. Then, it implies that the spin connection is totally determined by the tetrad configuration. Then, since $i\slashed{\nabla}$ is a self-adjoint operator \cite{rovelli} in Hilbert space with a scalar product given by
\bea  \int d^4x\sqrt{-g}\bar{\psi}(x)\phi(x),   \label{48}      \eea
which can be extended for our case since the generalized dual can be cast as  $\gdualn{\uplambda}=\bar{\psi}(x)$, with ${\psi}(x)=\frac{i\slashed{\nabla}\uplambda}{m}$ in terms of the Dirac dual, one can prove that the action in curved space-time is indeed Hermitian. Alternatively, one can also invoke the fact that since the first-order theory in curved space is evidently Hermitian, the path integration must keep this feature in its second derivative order version. \\
\indent On the other hand, in order to study quantum gravity, the metric is assumed to be a field operator. Then, the   path integral procedure becomes much harder. The effective adjoint would imply the use of a composite field in the effective action. It would involve field-dependent Jacobians that should be exponentiated to derive an effective model. Then, in this case, the first-order formulation presented here would be more appropriate.\\
\indent In order to summarize our last observations, we provide some additional consistency checks. The interaction part (with Higgs) is Hermitian already at the Lagrangian level, by just considering the definition of the dual.  In a curved space-time, the dual is generalized (we are considering c-number metric and geometric objects), but the proof of Hermiticity of the interaction Lagrangian is totally analogous to the flat space-time case. The Gaussian part can be shown to be Hermitian just at the action (discarding boundary term) level, considering the self-adjoint nature of the derivative operator in flat space-time case. As well as mentioned in \eqref{path}, the integration of the curved space-time version (considering a curved space-time in which the metric is A c-number, not a field, as previously mentioned ) of the first order action\footnote{built in terms of the Dirac dual and immediately Hermitian if one considers the fact that the covariant derivative contracted with the Dirac matrices times an imaginary factor is also self-adjoint in curved space time, see \eqref{48}}  would lead to the curved space-time version of \eqref{46}. Therefore, the path integral integration would not violate this property from the original equivalent model. Moreover, using \eqref{48}, the action in curved space-time can be proved to be Hermitian exactly as in the flat case. \\
\indent Considering the gravitational interaction, the geometric tensors are quantum fields. This point will be more consistently addressed in a forthcoming paper. However, the curved space-time version (fields, not c-numbers) of the equivalent first order formulation of \eqref{path}, is clearly Hermitian considering \eqref{48}.
Therefore, its integration is expected to be also Hermitian. However, since the geometry is now dynamical (not c-number anymore), the Elko dual would be a composite operator. It can lead to difficulties regarding Jacobians, etc in the path integration. Then, we concluded that, for the gravitational interaction, the first order (equivalent) version would be simpler to deal with. Then, analogously to the Dirac case, it would clearly lead to a Hermitian energy momentum tensor and corresponding interaction action.
.\\

\section{Concluding Remarks}\label{finalremarks}
In this work, we present a review of Elko spinors and the main recent developments --- in particular, Hermiticity. As is well known, Elko spinors are natural candidates to describe dark matter due to the intrinsic features of their construction. Consequently, these spinors and their associated quantum fields have been gaining considerable attention across various areas of physics. The approach we establish aims primarily to highlight the spinorial structure, the double Wigner degeneracy, and, consequently, the precise and correct definition of the dual structure providing consistence and physical meaning.
The new and correct adjoint structure allowed the definition of a Hermitian and renormalizable interaction with the Higgs. We showed that Elko inherently forbid standard minimal couplings with electromagnetic and non-Abelian fields. Alternative Hermitian couplings were proposed. Feynman rules for external Elko fields were derived, along with discussions on 1-loop radiative corrections and renormalizability. Tree-level Møller-like scattering was also analyzed, including explicit results for positive-definite squared amplitudes in polarized and unpolarized cases. Finally, we provided a first-order equivalent formulation by means of path-integral methods which is compatible with a straightforward introduction of a curved background suitable for coupling with gravity, the most fundamental interaction associated to dark-matter phenomenology.

\subsection*{Acknowledgment}
The authors express their gratitude to the editor Julio Marny Hoff da Silva for the kind and honorable invitation to contribute to this special issue.

\appendix


\begin{thebibliography}{40}

\bibitem{paper} G.B. de Gracia, et al. On Wigner Degeneracy in Elko theory: Hermiticity and Dark Matter, Physics of the Dark Universe {\bf 47}, 101774 (2025).


\bibitem{DM0}
G. Bertone \textit{et al.}, Particle dark matter: evidence, candidates and constraints, Physics Reports {\bf 405}, 5 (2005).

\bibitem{DM1}
N. Bozorgnia \textit{et al.}, Dark matter candidates and searches. Canadian Journal of Physics (2024).

\bibitem{DM2}
C. Balazs \textit{et al.}, A Primer on Dark Matter. arXiv preprint arXiv:2411.05062 (2024).

\bibitem{DM3}
X. Chen, and A. Loeb. Evolving Dark Energy or Evolving Dark Matter?. arXiv preprint arXiv:2505.02645 (2025).

\bibitem{DM4}
A. Chakraborty \textit{et al.}, Cosmological constraints on mass-varying dark matter. Physical Review D {\bf 111}, 6 (2025).

\bibitem{glens1}
M. Tajalli \textit{et al.}, SHARP--IX. The dense, low-mass perturbers in B1938+ 666 and J0946+ 1006: implications for cold and self-interacting dark matter. arXiv preprint arXiv:2505.07944 (2025).

\bibitem{glens2}
L. C. N., Santos \textit{et al.}, Black holes as gravitational mirrors. Universe {\bf 11}, 5 (2025)

\bibitem{glens3}
S. Hou \textit{et al.}, A Universal Analytic Model for Gravitational Lensing by Self-Interacting Dark Matter Halos. arXiv preprint arXiv:2502.14964 (2025).

\bibitem{glens4}
A. Tizfahm \textit{et al.}, Gravitational lensing in more realistic dark matter halo models. Physics of the Dark Universe {\bf 46},  101712 (2024).

\bibitem{glens5}
S. Vegetti \textit{et al.}, Strong gravitational lensing as a probe of dark matter. Space Science Reviews {\bf 220}, 5 (2024).

\bibitem{glens6}
M. Bílek, Peculiar dark matter halos inferred from gravitational lensing as a manifestation of modified gravity. Astronomy and Astrophysics {\bf 690}, A364 (2024).

\bibitem{glens7}
E. Seo \textit{et al.}, Inferring properties of dark galactic halos using strongly lensed gravitational waves. The Astrophysical Journal {\bf 966},1 (2024).

\bibitem{glens8}
A. Heavens, Weak lensing: Dark Matter, Dark Energy and Dark Gravity, Nucl.Phys.Proc.Suppl. {\bf 194}, 76 (2009).

\bibitem{cluster} G.Beck and M. Sarkis, Galaxy clusters in high definition: A dark matter search, Phys. Rev. D {\bf 107}, 023006 (2023).

\bibitem{curves} T. Ren \emph{et al.}, Reconciling the diversity and uniformity of galactic rotation curves with self-interacting dark matter, Phys. Rev. X {\bf 9}, 031020 (2019).

\bibitem{cdm} L. Perivolaropoulos, F. Skara, Challenges for $\Lambda$CDM: An update, New Astronomy Reviews, {\bf 95}, 101659 (2022).


\bibitem{Persic:1995ru}
M.~Persic, P.~Salucci and F.~Stel, The Universal rotation curve of spiral galaxies: 1. The Dark matter connection,
Mon. Not. Roy. Astron. Soc. \textbf{281}, 27 (1996).

\bibitem{Planck:2018vyg}
N.~Aghanim \textit{et al.} [Planck], Planck 2018 results. VI. Cosmological parameters, Astron. Astrophys. \textbf{641}, A6 (2020)

\bibitem{XENON:2022ltv}
E.~Aprile \textit{et al.} [XENON],
Search for New Physics in Electronic Recoil Data from XENONnT,
Phys. Rev. Lett. \textbf{129}, 16 (2022).

\bibitem{Sassi:2022njl}
S.~Sassi \textit{et al.}, Energy loss in low energy nuclear recoils in dark matter detector materials, Phys. Rev. D \textbf{106}, 6 (2022).

\bibitem{LZ:2022lsv}
J.~Aalbers \textit{et al.} [LZ], First Dark Matter Search Results from the LUX-ZEPLIN (LZ) Experiment,
Phys. Rev. Lett. \textbf{131}, 4 (2023). 

\bibitem{elkodark1}
J. M. Hoff da Silva, S. H. Pereira, Exact solutions to Elko spinors in spatially flat Friedmann-Robertson-Walker spacetimes, JCAP {\bf 03}, 009 (2014).

\bibitem{elkodark2}
A. Alves, \textit{et al.}, Searching for Elko dark matter spinors at the CERN LHC, Int. Jour. of Mod. Phys. A {\bf30}, 1550006 (2015).

\bibitem{elkodark3}
R. da Rocha \textit{et al.}, Exotic Dark Spinor Fields, JHEP {\bf 110}, 1104 (2011).

\bibitem{elkodark4}
R. da Rocha, J. M. Hoff da Silva, ELKO Spinor Fields: Lagrangians for Gravity derived from Supergravity, Int. J. Geom. Meth. Mod. Phys. {\bf 6} (2009).

\bibitem{elkodark5}
A. Alves \textit{et al.}, Constraining Elko Dark Matter at the LHC with Monophoton Events,  EPL {\bf 121}, 31001 (2018). 

\bibitem{elkodark6}
S. H. Pereira \textit{et al.}, $\Lambda(t)$ cosmology induced by a slowly varying Elko field, 	JCAP {\bf 01}, 055 (2017).


\bibitem{NPB2023} G.~B.~de Gracia, A.~A.~Nogueira and R.~da Rocha, Fermionic dark matter-photon quantum interaction: A mechanism for darkness, Nucl. Phys. B \textbf{992}, 116227  (2023).



\bibitem{elkosmology1}
S. H. Pereira \textit{et al.}, Evolution of the universe driven by a mass dimension one fermion field, EPL {\bf 120}, 3 (2017).

\bibitem{elkosmology2}
R. de C. Lima \textit{et al.}, Gravitational entropy of wormholes with exotic matter and in galactic halos, 	Int. J. Mod. Phys. D {\bf 29}, 2050015 (2020).

\bibitem{elkosmology3}
A. A. Escobal \textit{et al.}, Cosmological Constraints on Scalar Field Dark Matter,  Int. J. Mod. Phys. D {\bf 30},  15, 2150108 (2021).

\bibitem{elkosmology4}
S. H. Pereira \textit{et al.}, Dark matter from torsion in Friedmann cosmology, Eur. Phys. J. C {\bf 82}, 356 (2022).

\bibitem{elkosmology5}
L. C. T. Brito \textit{et al.}, Fermionic wave functions and Grassmann fields as possible sources of dark energy, Eur. Phys. J. C {\bf 82}, 821 (2022).






\bibitem{spinorsdm1}
A. Comech \textit{et al.}, Stable bi-frequency spinor modes as Dark Matter candidates. arXiv preprint arXiv:2501.04027 (2024).

\bibitem{spinorsdm2}
Y. Q. Gu, Nonlinear Spinors as the Candidate of Dark Matter. Open Access Library Journal, 4 (2017).

\bibitem{spinorsdm3}
A. B. Balakin and A. O. Efremova, Interaction of the axionic dark matter, dynamic aether, spinor and gravity fields as an origin of oscillations of the fermion effective mass. The European Physical Journal C {\bf 81}, 1 (2021).

\bibitem{Ahluwalia:2023slc}
D.~V.~Ahluwalia \textit{et al.}, Irreducible representations of the inhomogeneous Lorentz group with two-fold Wigner degeneracy, J. High Energ. Phys. {\bf} 75, 04 (2024).

\bibitem{dharamboson}
 D. V. Ahluwalia, Spin-half bosons with mass dimension three-half: Towards a resolution of the cosmological constant problem, Eur. Phys. Lett {\bf 131}, 41001 (2020)

\bibitem{dharamnpb} D.~V.~Ahluwalia, J.~M.~H.~da Silva and C.~Y.~Lee, Mass dimension one fields with Wigner degeneracy: A theory of dark matter, Nucl. Phys. B \textbf{987}, 116092 (2023).

\bibitem{rodo1} R. J. Bueno Rogerio, and G.B. de Gracia, Spinor Adjoints, Gauge Invariance and a new Road to Sterile Neutrinos, Physics of the Dark Universe, {\bf 47}, 101801 (2025).

\bibitem{rodo2}G. B. de Gracia, R.J. Bueno Rogerio, and L. Fabbri, Unraveling the Physical Meaning Behind Elko's Dual structure, Proc. R. Soc. A {\bf 480}, 20240349 (2024).

\bibitem{cheng} C.-Y. Lee \textit{et al.}, Mass dimension one fermions in FLRW space-time. Eur. Phys. J. C {\bf 85}, 267 (2025).

\bibitem{cosmo1}
D. Wang, Evidence for Dynamical Dark Matter. arXiv preprint arXiv:2504.21481 (2025).

\bibitem{cosmo2}
D. Friedan, A theory of the dark matter. arXiv preprint arXiv:2203.12405 (2022).


\bibitem{astrodm1}
F. Grippa, G. Lambiase, and T. K. Poddar, Searching for New Physics in an Ultradense Environment: A Review on Dark Matter Admixed Neutron Stars. Universe {\bf11}, 3 (2025).

\bibitem{astrodm2}
A. Khmelnitsky and V. Rubakov, Pulsar timing signal from ultralight scalar dark matter, J. Cosmol. Astropart. Phys. {\bf 02}, 019 (2014).

\bibitem{astrodm3}
J. A. Dror, and S. Verner. New Method for the Astrometric Direct Detection of Ultralight Dark Matter. Physical Review Letters {\bf 134}, 11 (2025).

\bibitem{astrodm4}
C. R. Arguelles \textit{et al.}, Fermionic dark matter: Physics, astrophysics, and cosmology. Universe {\bf 9}, 4 (2023).



\bibitem{wigner1} E.~P.~Wigner,  \emph{Unitary representations of the inhomogeneous Lorentz group including reflections}, in 
Group Theoretical Concepts and Methods in Elementary Particle Physics (Lectures of the Istanbul
Summer School of Theor. Phys.  (1962)), ed. F. G\"ursey, New York: Gordon and Breach, 1964.

\bibitem{wigner2}
E.~P.~Wigner, On Unitary Representations of the Inhomogeneous Lorentz Group,
Annals Math. \textbf{40} (1939) 149 [Nucl. Phys. Proc. Suppl. {\bf 6} (1989) 9 (Reprint)].

\bibitem{jcap}
D. V.Ahluwalia-Khalilova, and D. Grumiller, Spin-half fermions with mass dimension one: theory, phenomenology, and dark matter, Journal of Cosmology and Astroparticle Physics, {\bf 07}, 012 (2005).

\bibitem{speranca} L.~D.~Speran\c{c}a, An Identification of the Dirac Operator with the Parity Operator,
Int. J. Mod. Phys. D \textbf{23}, 14 (2014).

\bibitem{rodolfohidden}
R. J. Bueno Rogerio \textit{et al.}, A note on hidden classes in spinor classification, Proc. Roy. Soc. A \textbf{481}, 20240458 (2025).

\bibitem{rodolfotaka}
R. J. Bueno Rogerio \textit{et al.}, Revisiting Takahashi's inversion theorem in discrete symmetry-based dual frameworks, Phys. Lett. A \textbf{481}, 129028 (2023).


\bibitem{juliorogerio}
Hoff da Silva, J. M., Cavalcanti, R. T., ``Further investigation of mass dimension one fermionic duals'',\\
\textit{Phys. Lett. A} \textbf{383}, 1683 (2019).


\bibitem{spin} L.J. Boya and E.C.G. Sudarshan, The spin-statistics theorem in arbitrary dimensions, Int.J. Theor. Phys. {\bf 46}, 3285 (2007).

\bibitem{opticaltheo}
Goldberger, M. L.; Watson, K. M. Collision Theory. New York: Wiley, 1964.

\bibitem{rohr}
F. Rorlich, Quantum Electrodynamics of Charged Particles without Spin, Phys. Rev. {\bf{80}}, 666 (1950). 












\bibitem{daRocha:2011yr}
R.~da Rocha, A.~E.~Bernardini and J.~M.~Hoff da Silva, Exotic Dark Spinor Fields, J. High Energ. Phys. \textbf{04}, 110 (2011).

\bibitem{shapiro} I.L. Shapiro, Covariant derivative of fermions and all that, Universe. {\bf 8}, 586 (2022).

\bibitem{rovelli} G. Landi and C. Rovelli, General relativity in terms of Dirac eigenvalues, Phys. Rev. Lett. {\bf 78}, 3051 (1997).



















 









\end{thebibliography}
\end{document}